\documentclass[twocolumn]{aastex701}

\begin{document}

\title{Orbital Architecture of Super-Earth--Cold-Jupiter Systems from Transit Duration Ratios}

\author[orcid=0009-0001-0736-7070,sname='Lassmann']{Tom Lassmann}
\affiliation{Institute for Astronomy, University of Hawai`i, 2680 Woodlawn Drive, Honolulu, HI 96822, USA}
\affiliation{University of Bremen, Department of Physics and Electrical Engineering, Bremen, Germany}
\affiliation{Boston University, Department of Astronomy, Boston, MA, USA}
\email{lassmann@bu.edu}

\author[orcid=0009-0004-9078-5987]{Joseph Hand}
\affiliation{Institute for Astronomy, University of Hawai`i, 2680 Woodlawn Drive, Honolulu, HI 96822, USA}
\email{jhand@hawaii.edu}

\author[orcid=0000-0002-8958-0683, sname='Dai']{Fei Dai} 
\affiliation{Institute for Astronomy, University of Hawai`i, 2680 Woodlawn Drive, Honolulu, HI 96822, USA}
\email{fdai@hawaii.edu}

\author[orcid=0000-0002-7563-7618, sname='Tang']{Jiaxin Tang} 
\affiliation{Tsinghua University, Department of Astronomy, Beijing, China}
\email{tangjx22@mails.tsinghua.edu.cn}  

\author[orcid=0000-0002-6937-9034, sname='Wang']{Sharon Wang} 
\affiliation{Tsinghua University, Department of Astronomy, Beijing, China}
\email{sharonw@tsinghua.edu.cn}  

\author[orcid=0000-0003-0707-3213,sname='Chen']{Di-Chang Chen}
\affiliation{School of Physics and Astronomy, Sun Yat-sen University, Zhuhai 519082, China}
\affiliation{CSST Science Center for the Guangdong-Hong Kong-Macau Great Bay Area, Sun Yat-sen University, Zhuhai 519082, China}
\email{chendch28@mail.sysu.edu.cn}

\begin{abstract}

Super-Earths are commonly found with cold Jupiters, yet it is unclear whether these Super-Earth--Cold-Jupiter (SE-CJ) systems differ architecturally from other Kepler-like systems (Non-SE-CJ). One might expect the CJs to excite eccentricities and mutual inclinations among the SEs. We use two metrics: the normalized transit duration $T/T_0$, set by the host-star density, and the pairwise duration ratio $\xi$ between neighboring transiting planets. We find no population-level difference in $T/T_0$ between the samples. Forward modeling of $\xi$, which is more directly sensitive to mutual inclinations because it isolates the impact-parameter contrast, finds no evidence that SE-CJ systems are more mutually inclined than Non-SE-CJ systems, and hints they could be even flatter. We infer a Rayleigh dispersion in mutual inclination of $\delta \approx 1.0 ^{\circ}$, or a characteristic eccentricity of $e \approx 0.035$ (assuming equipartition), compared with $\delta \approx 1.5 ^{\circ}$ and $e \approx 0.05$ for Non-SE-CJ systems.  Many SE-CJ systems therefore retain nearly coplanar, low-eccentricity architectures despite hosting distant giants. We further show that secular coupling among the inner Super-Earths usually dominates over forcing from the cold Jupiter, leaving the inner planets resistant to excitation. Our analysis is limited to SE-CJ systems with multiple transiting SEs: systems in which the CJ has strongly excited the inner architecture are expected to present as single-transiting, and cannot be studied with the $\xi$ metric. Another caveat of our result is that undetected cold Jupiters in the comparison sample could dilute the measured $\xi$ distribution.



\end{abstract}

\keywords{\uat{Exoplanets}{498} --- \uat{Exoplanet astronomy}{486} --- \uat{Super Earths}{1655} --- \uat{Exoplanet systems}{484} --- \uat{Exoplanet dynamics}{490} --- \uat{Exoplanet detection methods}{489} --- \uat{Extrasolar gaseous giant planets}{509} --- \uat{Inclination}{780} --- \uat{Eccentricity}{441} --- \uat{Orbital elements}{1177} --- \uat{Transits}{1711} --- \uat{Transit photometry}{1709} }


\section{Introduction} \label{sec:intro}
Compact ($<1\,\mathrm{AU}$) planetary systems containing multiple transiting Super-Earths/sub-Neptunes\footnote{In this work, we use the terms `Super-Earth' and `sub-Neptune' interchangeably as an all-catching term for planets with $R_p<5R_\oplus$.}  are common in our Galaxy, as demonstrated by the \textit{Kepler} mission \citep{Borucki,Petigura2013KeplerOccurence,Zhu2018}. These `Kepler-like' systems are generally dynamically cold, with low mutual inclinations \citep[typically $1$--$2.2^\circ$ see e.g. ][]{Fabrycky2014,Lissauer2023KeplerOverview,Fang} and low orbital eccentricities \citep[e.g., $e\simeq0.05\pm0.01$;][]{Gilbert2025EccentricityValues,Xie,VanEylen,He2020}. 

Cold Jupiters (CJ) are frequently found in systems that host inner Super-Earths (SE): stars with inner SEs have a $\sim20$--$40\%$ probability of harboring a CJ at $\gtrsim1\,\mathrm{AU}$ \citep[e.g.,][]{Zhu2018SECJoccurence,Bryan2019SECJexcess,Rosenthal2022CLS_SECJ,Zhu2024SECJmetallicity,Bryan2024,Liu2026TESS_SECJOccurence}, about a factor of 2--4 higher than the field occurrence rate of CJs ($\sim10\%$; \citealt{Fernandes2019GiantOccurence,Fulton2021CLSGiantOccurence,VanZandt}). The CJ excess is more pronounced around metal-rich hosts \citep{Zhu2024SECJmetallicity,Bryan2024,Bonomo2025}; and it appears stronger for gas-rich sub-Neptunes rather than for rocky Super-Earths \citep{Chen2026SupNeptuneCJ}. Together, these observations seem to suggest a formation pathway in which metal-rich protoplanetary disks facilitate the formation of both gas-rich sub-Neptunes and giant planets. The early formation of SEs may also enable them to migrate inward to the $\sim0.1\,$AU orbits where they reside today \citep{Zhu2024SECJmetallicity,Schlecker2020SECJarchitecureLink,Bitsch2020EccentricGiants,Ogihara}; whereas slower Type-II migration of CJs often keeps them at $>1$\,AU \citep{Kley}.


The frequent coexistence of SEs and CJs raises a central dynamical question: do outer giant planets dynamically excite inner systems on a population level, or can tightly packed inner planetary chains resist such perturbations and remain dynamically cold? 


The Solar System provides a useful point of comparison: our giant planets play a central role in shaping the eccentricities and inclinations of the terrestrial planets \citep{Murray1999SolarSystemDynamics}, and ultimately the Solar System's long-term stability \citep{Laskar2009,Lithwick2011}. SE-CJ systems may or may not resemble the Solar System. On one hand, tightly packed inner systems can be secularly or resonantly coupled with each other strongly, causing them to undergo coordinated nodal precession that suppresses mutual inclination excitation even in the presence of a modestly inclined CJ \citep{Rodet2021InclinedCJPertubation,Spalding}. On the other hand, if the outer giants are close enough and dynamically hot themselves, they can torque inner planetary systems through secular interactions, exciting mutual inclinations and eccentricities \citep{Lai2017CJPerturbInclination,Pu2021ScatteringCJInfluence,Masuda2020CJinclinationPertubator,Guo2026Impact}, with Kepler-139 as a candidate observational example \citep{Lammers}. The surviving SEs may be left on eccentric and mutually inclined orbits which can reduce the observed transit multiplicity (number of observed transiting planets in a system). This anti-correlation between mutual inclinations and transit multiplicity has been reported in several previous works \citep{Pu2021ScatteringCJInfluence,Zhu2018,Millholland_TDV}.

In this paper, we empirically tested whether the orbital architectures of SE-CJ systems differ from the broader Kepler-like systems (Non-SE-CJ), using two transit duration metrics:

\begin{enumerate}
    \item  {\bf Normalized Transit Duration $T/T_0$}: $T_0$ is the expected duration for a central, circular orbit \citep{Fabrycky2014,Xie}. Because $T_0$ is defined at $b=e=0$, deviations of $T/T_0$ from unity reflect a combination of non-zero impact parameter $b$ and eccentricity
argument $e\sin\omega$. At the population level, a systematic shift between SE-CJ and Non-SE-CJ would signal different mutual inclination and/or eccentricity distribution. One can compute $T/T_0$ for all transiting planets (including singles) as long as the stellar mean density has been measured precisely \citep{Winn2014TransitsOccultations},
\begin{equation}
    T_0 = \left(\frac{3P}{\pi^2 G \rho_\star}\right)^{1/3},
    \qquad
    \rho_\star = \frac{M_\star}{\frac{4}{3}\pi R_\star^3}.
    \label{eq:T0}
\end{equation}
    \item  {\bf Pairwise Transit Duration Ratios $\xi$}: i.e. the transit duration ratio of neighboring transiting planets in the same system:
\begin{equation}
    \xi \equiv \frac{T_{\rm in}/P_{\rm in}^{1/3}}
                    {T_{\rm out}/P_{\rm out}^{1/3}},
    \qquad x \equiv \log_{10}\xi,
    \label{eq:xi}
\end{equation}
which again depends on the impact parameter (mutual inclination) and the eccentricity vector of the transiting planets. Previous works have utilized this metric to constrain the population-level eccentricity distribution of the Kepler-like planets \citep{Fabrycky2014}. 
\end{enumerate}


The paper is organized as follows. Section~\ref{sec:data} describes the SE-CJ and Non-SE-CJ samples. Sections~\ref{sec:tt0} and \ref{sec:xi} present the analyses of $T/T_0$ and $\xi$, respectively. Section~\ref{sec:discussion} discusses the implications of our findings. Section \ref{sec:conclusions} is a summary of this paper.

\section{Sample Construction} \label{sec:data}

We define a cold Jupiter (CJ) as any confirmed or candidate planet satisfying $P>100\,\mathrm{d}$ and $R_p>5\,R_\oplus$; for non-transiting RV detections lacking a measured radius, we convert the published mass to an approximate radius using \texttt{Forecaster} \citep{Chen2016Forecaster} before applying this criterion. A Super-Earth (SE) is any planet with $P<100\,\mathrm{d}$ and $R_p<5\,R_\oplus$. An SE-CJ system contains at least one of each. This definition is intentionally broad, encompassing both rocky Super-Earths and volatile-rich sub-Neptunes, as well as Saturn-sized giant planets to increase the overall sample size. 


\subsection{CKS sample for \texorpdfstring{$T/T_0$}{T/T0} Analysis}

Since the analysis of the $T/T_0$ metric relies on precise stellar density, we used the homogeneously-constructed \textit{California Kepler Survey} (CKS) sample \citep{Petigura2017CKS_I, Johnson2017CKS_II}. We cross-matched the CKS sample with Kepler DR25 for transit parameters \citep{Thompson}. Before filtering, the catalog contains 2025 entries across 1279 systems, of which 1508 are confirmed or candidate planets. We retained only confirmed and candidate planets (removing false positives and undispositioned signals). We kept only systems with Kepler magnitude $Kp<16.2$, stellar surface gravity $3.3<\log g<10.0$, and effective temperature $3000\,\mathrm{K}<T_{\rm eff}<7000\,\mathrm{K}$, restricting the sample to well-characterized Sun-like host stars. After these cuts, the sample contains 1482 planets in 891 systems, including 514 single-transiting systems and 377 multi-transiting systems. The planet sample comprises 1226 SEs, 69 CJs, and 16 SE-CJ systems. We additionally include 12 single-transiting SEs in known SE-CJ systems that were
not observed by CKS. For these systems, the stellar mean densities have been measured precisely from high-resolution spectroscopy almost identical to the method employed in the CKS analysis \citep[e.g.][]{Weiss}. Our conclusions are unchanged if this supplementary sample is excluded.


\subsection{Literature Sample for \texorpdfstring{$\xi$}{xi} Analysis}

\begin{figure*}[t]
    \makebox[\textwidth][c]{%
        \includegraphics[width=1.\textwidth]{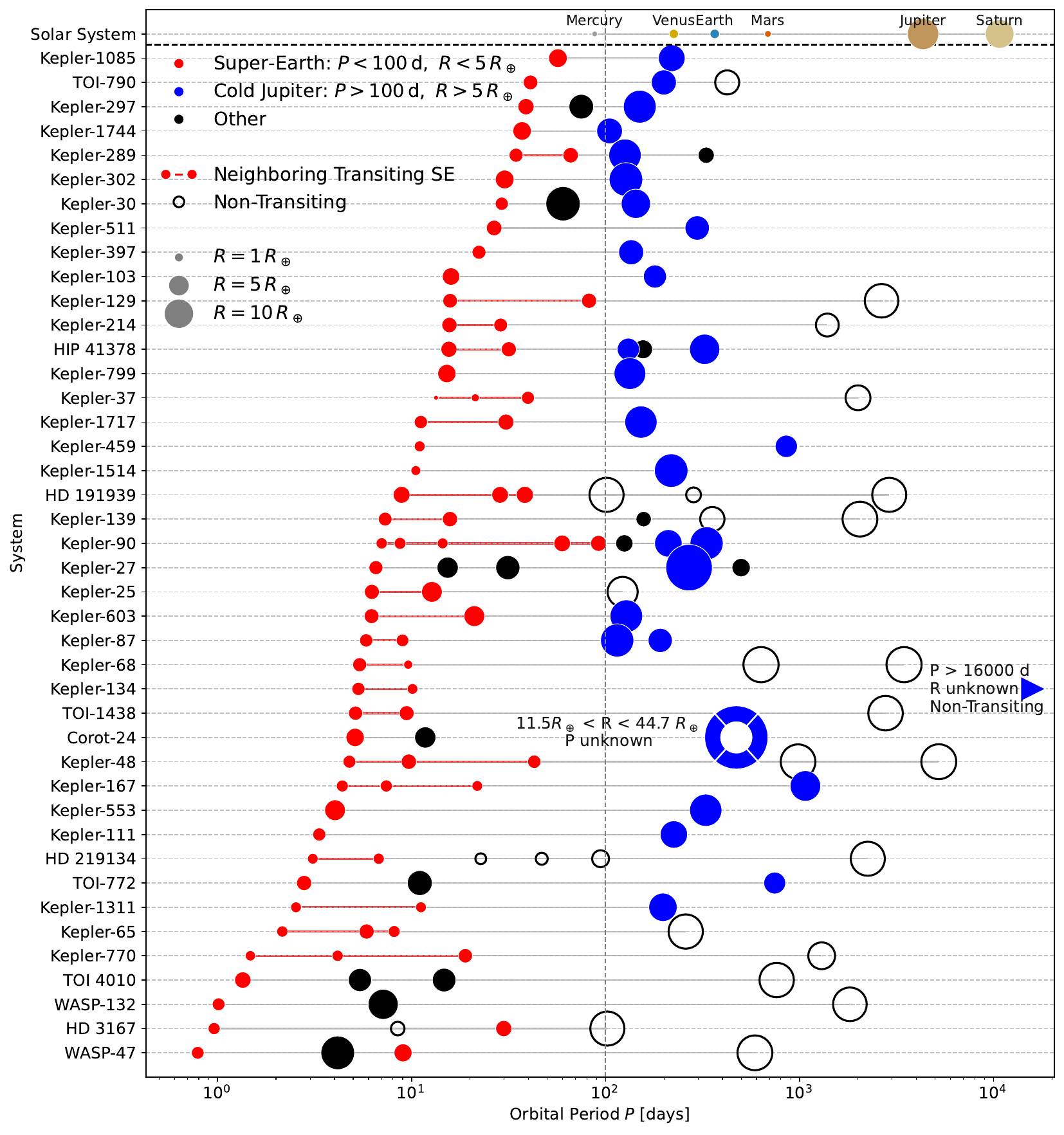}
    }
    \caption{Orbital architecture of the multi-transiting SE-CJ systems. Planets are categorized into SEs (red), CJs (blue) and Others (black). Red lines connect neighboring transiting SE-SE pairs included in the duration-ratio analysis. Non-Transiting planets are marked with an open circle (no fill). For non-transiting CJs lacking a measured radius, planet radii are estimated from published masses using \texttt{Forecaster} \citep{Chen2016Forecaster}. Planets with some unknown parameters are specially annotated in the plot. Solar System planets within range 
    were added and annotated in the top row for comparison. Solar System parameters were taken from \citet{Seidelmann2007SolarValues1, Seidelmann1992SolarValues2}.}
    \label{Results.Architecture.Image}
\end{figure*}

Since $\xi$ analysis is agnostic to the properties of the host star, we assembled a larger sample of all confirmed multi-transiting SE-CJ systems from the Exoplanet Follow-up Observing Program
(ExoFOP, see Acknowledgments), NASA Exoplanet Archive\footnote{https://exoplanetarchive.ipac.caltech.edu} and the Kepler DR25 catalog \citep{Thompson}. This effort adds 75 confirmed and candidate transiting planets to the CKS SE-CJ sample. The complete list and architecture of our 42 multi-transiting SE-CJ sample are shown in Figure~\ref{Results.Architecture.Image}. The Non-SE-CJ comparison sample comprises all remaining systems in the CKS sample, i.e.\ those with no detected CJ meeting the criteria above. Combined with the CKS sample of Section~\ref{sec:data}, the full planet sample used for the $T/T_0$ analysis comprises 1568 planets.

Two-sample Kolmogorov-Smirnov tests confirm that the SE-CJ and Non-SE-CJ samples are statistically indistinguishable in their planetary ($R_p$, $P$) and stellar ($T_{\rm eff}$, $M_\star$, $R_\star$) properties (all KS $p>0.1$), with only host-star metallicity directionally higher for SE-CJ systems ($p=0.08$) as expected from the SE-CJ-metallicity correlation \citep{Zhu2024SECJmetallicity, Bryan2024, Bonomo2025}. 
Of our 42 SE-CJ systems, 18 have CJs detected only by radial velocity or transit-timing variations; all 42 are eligible for the $\xi$ analysis, which uses only the inner SE-SE transit pairs; the CJ detection method is therefore irrelevant to $\xi$. Of these, 20 systems contribute at least one adjacent SE-SE pair.\footnote{%
TOI-790.03 is transiting but has only a single observed transit with insufficient parameter constraints. KOI-693.03 and KOI-1108.04 transit
but lack the parameters required for our analysis
\citep{Kawahara2019DataCJsMissingTValue}.}

\section{Analysis of Normalized transit duration \texorpdfstring{$T/T_0$}{T/T0}}
\label{sec:tt0}

\begin{figure*}[t!]
    \centering
    \includegraphics[width=\textwidth]{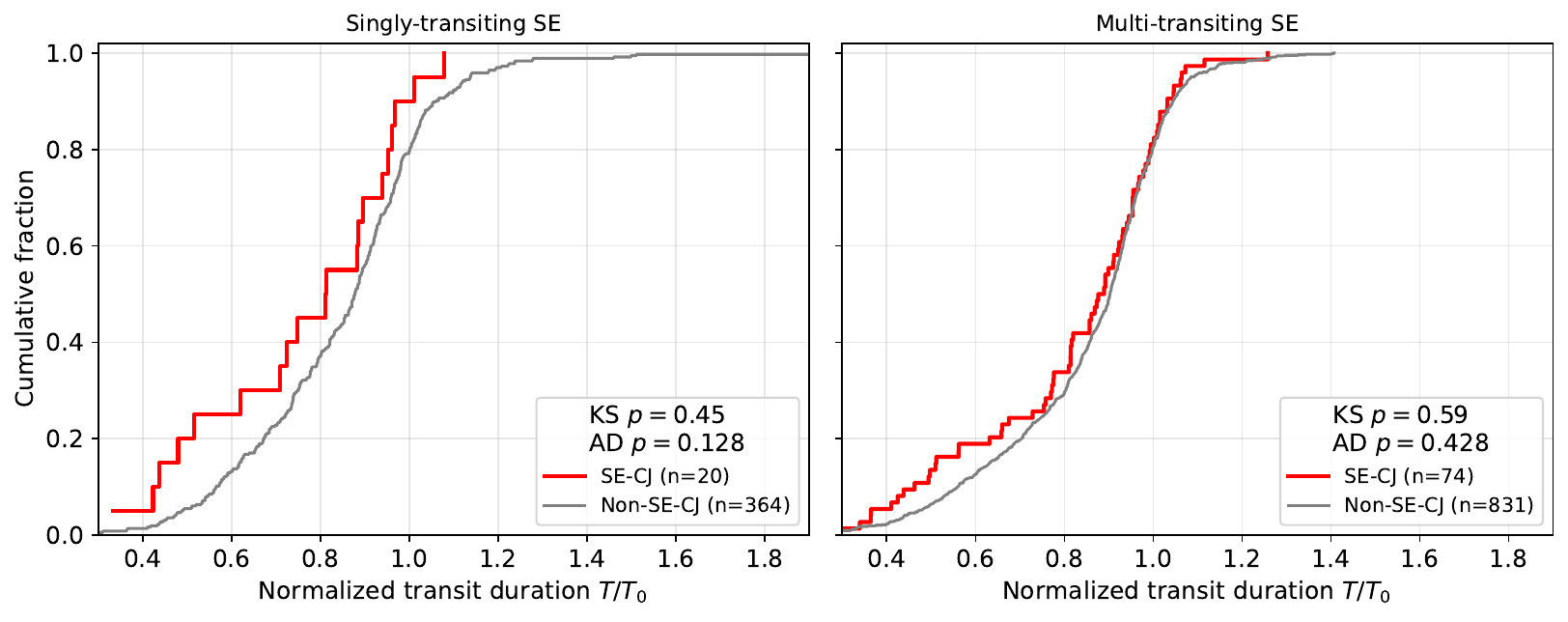}
    \caption{ECDF of the normalized transit duration $T/T_0$ for
    Super-Earth planets split by transit multiplicity.
    \textit{Left:} singly-transiting SE in SE-CJ vs.\ Non-SE-CJ systems
    ($N_{\rm SECJ}=20$, $N_{\rm Non}=364$; KS $p=0.45$, AD $p=0.13$).
    \textit{Right:} multi-transiting SE in SE-CJ vs.\ Non-SE-CJ systems
    ($N_{\rm SECJ}=74$, $N_{\rm Non}=831$; KS $p=0.59$, AD $p=0.43$).
    No significant difference is detected in either subset.}
    \label{fig:tt0}
\end{figure*}

We computed the normalized transit duration $T/T_0$ using Eqn. \ref{eq:T0} and the median values of the stellar and planetary parameters assembled in Section \ref{sec:data}. Figure~\ref{fig:tt0} compares the $T/T_0$ distributions for SE planets in SE-CJ and Non-SE-CJ systems. A two-sample Kolmogorov-Smirnov test implemented in \texttt{scipy} 
yields $p=0.20$. Thus no significant difference is detected between the SE-CJ and Non-SE-CJ systems. In other words, the existing sample does not support more dynamically stirred orbital architectures in SE-CJ systems.

We further split the sample by transit multiplicity ($N_{\rm tra}=1,2,3,\geq 4$) in anticipation that systems with lower transit multiplicity may show higher mutual inclination \citep{Zhu2018,Millholland_TDV}.  We found KS-test $p$-values above $0.3$ in every subset except one across singly-transiting, multi-transiting, and individual transit-multiplicity bins ($N_{\rm tra}=1,2,3,\geq 4$) between the SE-CJ and Non-SE-CJ systems; only the $N_{\rm tra}\geq4$ bin falls below $p=0.05$ ($p=0.017$, $N_{\rm SE-CJ}=15$), which is not significant after accounting for the four bins tested. We therefore found no evidence for a population-level eccentricity contrast between SE-CJ and Non-SE-CJ systems based on $T/T_0$ alone. We do note that within the SE-CJ sample, single-transiting SEs show a lower $T/T_0$ than SE-CJ with multiple-transiting SEs (with median at $0.81$ vs.\ $0.88$), directionally consistent with single-SE-CJ systems being more dynamically excited \citep{Huang2017hotSEfromScattering}. A similar trend has been noted in previous work for the Non-SE-CJ systems \citep{Xie}.  However, this within-SE-CJ difference is not statistically significant given the small sample size of singly-transiting SE-CJ systems ($n=20$, KS $p=0.49$).

\section{Analysis of pairwise transit duration ratios \texorpdfstring{$\xi$}{xi}}
\label{sec:xi}

We next examined the more sensitive pairwise transit duration ratios $\xi$. Within each system, the planets were sorted by orbital periods. $\xi$ was then computed between neighboring pairs of planets where both planets
satisfy the SE criteria ($P<100\,\mathrm{d}$, $R_p<5\,R_\oplus$) and both have $\mathrm{S/N}>7.1$\footnote{This threshold was applied to every planet using its catalog
signal-to-noise ratio. Two SE-CJ planets have no entry in the cumulative KOI
table; for these we adopt $\mathrm{S/N}=8.7$ for Kepler-90\,i
\citep{Shallue2018Kep90} and $\mathrm{S/N}=20.5$ for Kepler-289\,d, scaled
from the other planets in the same system. For planets in bright-star K2/TESS systems
(HD~191939, HD~219134, HD~3167, HIP~41378, TOI-1438), and for Kepler-289\,d, the Kepler
$\mathrm{S/N}>7.1$ pipeline threshold does not apply, and we instead adopt the
purely geometric detectability limit $b_{\rm max}=1+k$; the S/N value therefore
does not enter the forward model for these planets.} (the Kepler detection threshold; \citealt{Jenkins2010KeplerPipeline}; \citealt{Christiansen2012CDPP}; \citealt{Fabrycky2014}). The log-duration ratio $x=\log_{10}\xi$ was trimmed to $|x| < 2.5$ to avoid sensitivity to outliers caused by poorly constrained grazing orbits. We did not impose any direct cut on impact parameter $b$: the S/N$>7.1$ requirement already rejects grazing transits that contribute negligible duration information, and the $|x|<2.5$ trim removes the residual high-$b$ outliers in $\xi$ space. We additionally require that both planets in a pair have a fractional
transit-duration uncertainty below $10\%$, applied identically to both
populations, since poorly constrained durations inject scatter into $\xi$ that
the forward model would otherwise absorb into $\delta$. The final sample
contained $N_{\rm SE-CJ} = 26$ pairs and $N_{\rm Non-SE-CJ} = 448$ pairs.


\subsection{Forward model of \texorpdfstring{$\xi$}{xi}}
\label{sec:forwardmodel}

We employed a model similar to that of \citet{Fabrycky2014}, parameterized by two population-level quantities: $\delta$ (the Rayleigh dispersion of the mutual inclination distribution) and $\kappa$ (the eccentricity-inclination scaling, $\sigma_e = \kappa\delta_{\rm rad}$, where $\delta_{\rm rad}$ is $\delta$ in radians). The eccentricity and inclination dispersion are linked because dynamical processes often excite both simultaneously \citep{Murray1999SolarSystemDynamics}. The same model was applied to both SE-CJ and Non-SE-CJ populations.

For each observed pair of transiting planets, we generated Monte Carlo samples of $\xi$ in three steps:

\begin{enumerate}

\item \textbf{Eccentricities}
We drew the eccentricity-vector components $(e\cos\omega, e\sin\omega)$ independently from $\mathcal{N}(0,\sigma_e^2)$ for both the inner and
outer planet, giving a Rayleigh eccentricity distribution with scale $\sigma_e = \kappa\delta_{\rm rad}$. Draws with $e\geq 1$ or with inner apocenter exceeding outer pericenter ($a_{\rm in}(1+e_{\rm in}) \geq a_{\rm out}(1-e_{\rm out})$) were rejected.

\item \textbf{Outer impact parameter}
The outer planet's impact parameter $b_{\rm out}$ was drawn uniformly in $[0, b_{\rm out,max}]$, where $b_{\rm out,max}$ is set by the Kepler
detectability condition $\mathrm{S/N}_{\rm model}\geq 7.1$, with:
\begin{equation}
\mathrm{S/N}_{\rm model} = \mathrm{S/N}_{\rm obs}\sqrt{T_{\rm model}/T_{\rm obs}}
\end{equation}

\item \textbf{Inner impact parameter}
The inner planet's impact parameter was drawn from
\begin{equation}
\begin{array}{cc}
b_{\rm in} \sim \mathcal{N}\!\left(
        b_{\rm out}\,\frac{a_{\rm in}/R_\star}{a_{\rm out}/R_\star},\;
        \sigma_b^2
    \right),\\
\sigma_b = \delta_{\rm rad}\,\frac{a_{\rm in}/R_\star}{1+R_{p,\rm in}/R_\star},
\end{array}
    \label{eq:bin_draw}
\end{equation}
following the geometry of \citet{Fabrycky2014}. Draws with $|b_{\rm in}|>b_{\rm in,max}$ (the inner detectability limit) were rejected. 

\end{enumerate}

Transit durations were then computed from the standard expression
\citep{Winn2014TransitsOccultations}:
\begin{equation}
\begin{array}{cc}
    T = \frac{P}{\pi}\arcsin\!\left(
        \frac{R_\star}{a}\,
        \frac{\sqrt{(1+k)^2-b^2}}{\sin i_{\rm los}}
    \right)
    \cdot\,g(e,\omega),
    \\
    g = \frac{\sqrt{1-e^2}}{1+e\sin\omega},
    \label{eq:Tdur}
\end{array}
\end{equation}
where $k=R_p/R_\star$ and $i_{\rm los}$ was obtained from $b$ by inverting the standard impact-parameter relation. Duration noise was added using each planet's catalog fractional duration
uncertainty $f_\sigma$, as
$\sigma_T = f_\sigma\,T_{\rm obs}\sqrt{T_{\rm model}/T_{\rm obs}}$; this is the
same quantity to which the duration-quality cut of Section~\ref{sec:xi} is
applied. Substituting the photometric form
$\sigma_T = T\sqrt{2k}/\mathrm{S/N}_{\rm model}$ of \citet{Fabrycky2014} shifts
$\delta_{\rm SE-CJ}$ by less than $0.05^\circ$. Each pair contributed 1000 accepted Monte Carlo realizations to the simulated $x$ distribution.

\subsection{Likelihood grid and Bayesian inference}
\label{sec:likelihood}

We evaluated a binned multinomial log-likelihood on a uniform grid with $0.01^\circ$ steps in $\delta \in [0.01^\circ, 4.00^\circ]$ and $0.1$ steps in $\kappa \in [0, 5]$. For posterior inference we restrict to $\delta \geq 0.2^\circ$: typical Kepler transit-duration uncertainties of a few percent translate, for our $a/R_\star \sim 20$ pairs, into inclination-equivalent noise of order $0.1$--$0.2^\circ$, below which the forward model is dominated by photometric noise rather than the inclination signal. The Jeffreys prior $p(\delta) \propto 1/\delta$ would otherwise concentrate posterior mass at the grid edge purely as a prior artifact. We confirmed that posterior medians are insensitive to the precise floor in the range $0.1^\circ$--$0.3^\circ$. The observed and simulated $x = \log_{10}\xi$ distributions were each divided into equal bins of width $\Delta\log_{10}\xi = 0.037$ spanning $[-1.5, 1.5]$. We tested a range of bin widths. For larger widths the inferred $\delta$ is unstable, varying by more than $20\%$ depending
on where the bin edges happen to fall. This sensitivity decreases steadily as
the bins are narrowed and becomes negligible once the width drops below about
$0.05$, with the inferred values converging. We therefore adopt
$\Delta\log_{10}\xi = 0.037$, the coarsest width in the converged regime. We
note that this is not the width that minimises $\delta_{\rm SE-CJ}$, but rather the most stable estimator choice. With the observed bin counts $\{n_b\}$ and the simulated bin probabilities $\{p_b(\delta, \kappa)\}$ derived from the forward model, the log-likelihood is
\begin{equation}
    \ln\mathcal{L}(\delta,\kappa) = \sum_b n_b \ln p_b(\delta,\kappa).
    \label{eq:loglike}
\end{equation}
Empty bins were assigned an infinitesimal value to avoid $\ln 0$. To suppress Monte Carlo noise in the likelihood surface, we ran the simulator independently with 10 random seeds and averaged the resulting likelihoods at each grid point (not the log-likelihoods), computed in numerically-stable log space as:
\begin{equation}
\ln\mathcal{L}(\delta,\kappa) = -\ln N_{\rm seed} + \ln\!\sum_{s=1}^{N_{\rm seed}} e^{\ln\mathcal{L}_s(\delta,\kappa)}.
\end{equation}

We adopted a log-uniform (Jeffreys) prior $p(\delta)\propto1/\delta$, appropriate for a scale parameter, and a uniform prior on $\kappa$.
Posteriors were obtained by direct normalization,
$p(\delta,\kappa\mid\mathrm{data})\propto
\mathcal{L}(\delta,\kappa)\,p(\delta)\,p(\kappa)$.
We verified that the main result was robust to this prior choice: replacing the Jeffreys prior with a flat prior yielded qualitatively similar results. As a further robustness check, we performed a system-level leave-one-out
test, removing all pairs contributed by each of the 20 systems in the SE-CJ
sample in turn and re-inferring $\delta_{\rm SE-CJ}$. The largest single-system
effect comes from Kepler-139, discussed separately below; excluding that system,
the posterior median shifts by at most $-0.22^\circ$ and $+0.19^\circ$. No single
system drives the result, and under every exclusion $\delta_{\rm SE-CJ}$ remains
below the Non-SE-CJ value. We also repeated the fit with the duration-quality cut of
Section~\ref{sec:xi} relaxed, applying the same threshold to both populations.
Raising it to $20\%$ admits three further SE-CJ pairs but no additional systems,
giving 29 pairs across the same 20 systems. Removing the cut entirely retains
all 32 SE-CJ pairs across 23 systems and gives
$\delta_{\rm SE-CJ}=1.60^{+0.90}_{-0.55}{}^\circ$ against
$\delta_{\rm Non-SE-CJ}=1.54^{+0.18}_{-0.15}{}^\circ$, at which point the two
populations are no longer distinguishable
($P(\delta_{\rm SE-CJ}>\delta_{\rm Non-SE-CJ})=0.54$). The pairs removed by the
cut have fractional duration uncertainties between $11\%$ and $36\%$, large
enough that their scatter is absorbed into the inferred inclination dispersion.
We therefore regard the cut as necessary for the comparison to be meaningful,
and apply it identically to both populations. \\ One system, Kepler-139, warrants individual attention. It contributes a
single pair, Kepler-139\,d and b, whose log duration ratio, $x=-0.44$, is the
largest deviation from unity in the SE-CJ sample; every other pair lies within
$|x|\leq0.25$, with a median $|x|$ of $0.05$. Excluding Kepler-139 lowers the
shared-$\kappa$ posterior median from $\delta_{\rm SE-CJ}=1.00^{+0.59}_{-0.37}{}^\circ$
to $0.67^{+0.42}_{-0.27}{}^\circ$, moves the separate-fit value from $0.86^\circ$
to $0.59^\circ$, and lowers $P(\delta_{\rm SE-CJ}>\delta_{\rm Non-SE-CJ})$ from
$0.204$ to $0.059$ ($0.115$ to $0.028$ for the separate fits).
There is reason to treat this particular pair with caution. Reproducing its
observed duration ratio requires an extreme geometry in every case. In a common
orbital plane with circular orbits the inner planet has the smaller impact
parameter, $b_{\rm in}/b_{\rm out}=(a_{\rm in}/R_\star)/(a_{\rm out}/R_\star)=0.60$,
so the outer planet grazes first and $\xi$ cannot fall below $0.99$, leaving the
observed $\xi=0.36$ unreachable. Allowing eccentricity admits coplanar
solutions, but only ones in which the inner planet transits within
$0.001\,R_\star$ of the grazing limit. Dropping coplanarity, the inner impact
parameter still never falls below $0.63$ across all combinations of inclination
and eccentricity that reproduce the ratio while keeping both planets in transit
and the orbits non-crossing, and reaching even that value requires
$e_{\rm in}\approx0.65$.
Independently, \citet{Lammers} find from a joint transit-timing and
radial-velocity analysis that Kepler-139\,d and b are the most tightly coupled
pair in this system, remaining mutually aligned to well below a degree, and
their analysis adopts rather than re-fits the catalogued transit-shape
parameters. We therefore flag Kepler-139 as warranting a targeted re-analysis
of its transit photometry. We nonetheless retain it in all quoted results, since
it meets every selection criterion and its exclusion would only strengthen our
conclusion. Because we fit a single Rayleigh distribution to all pairs, retaining this system
raises the inferred $\delta_{\rm SE-CJ}$ and therefore weakens the contrast we
report.


The parameter $\kappa = \sigma_e/\sigma_i$ encodes how strongly eccentricity and inclination are coupled in the underlying dynamics. We experimented with allowing $\kappa$ to differ or be held constant (\emph{shared-$\kappa$} model) between the SE-CJ and Non-SE-CJ model.
In the \emph{shared-$\kappa$} model, the joint three-dimensional log-likelihood
$\mathcal{L}_{\rm tot}\equiv \mathcal{L}_{\rm tot}(\delta_{\rm SE-CJ}, \delta_{\rm Non-SE-CJ}, \kappa)$ is:
\begin{eqnarray}
    \ln\mathcal{L}_{\rm tot} &=&
    \ln\mathcal{L}_{\rm SE-CJ}(\delta_{\rm SE-CJ},\kappa) \nonumber \\
    && +\ \ln\mathcal{L}_{\rm Non-SE-CJ}(\delta_{\rm Non-SE-CJ},\kappa),
\end{eqnarray}
where both populations share a single $\kappa$ but have independent
$\delta$ values. We normalize the joint posterior
$p(\delta_{\rm SE-CJ},\delta_{\rm Non-SE-CJ},\kappa\mid\mathrm{data})\propto \mathcal{L}_{\rm tot}\,p(\delta_{\rm SE-CJ})\,p(\delta_{\rm Non-SE-CJ})\,p(\kappa)$
by summing over all grid points. We report $P(\delta_{\rm SE-CJ}>\delta_{\rm Non-SE-CJ})$, the posterior probability that the SE-CJ systems are the \emph{more} dispersed of the two, which is given by:
\begin{eqnarray}
    P &=& \int\!\!\int_{\mathcal{D}}
    p(\delta_{\rm SE-CJ},\delta_{\rm Non-SE-CJ}\mid\mathrm{data}) \nonumber \\
    && \qquad \mathrm{d}\delta_{\rm SE-CJ}\,\mathrm{d}\delta_{\rm Non-SE-CJ},
\end{eqnarray}
where $\mathcal{D}$ is the region $\delta_{\rm SE-CJ}>\delta_{\rm Non-SE-CJ}$ and $p(\delta_{\rm SE-CJ},\delta_{\rm Non-SE-CJ}\mid\mathrm{data})$ is the joint posterior marginalized over $\kappa$.

We report $\delta$ values as posterior medians with 68\% credible intervals. The Non-SE-CJ posterior peaks near $\delta\approx1.5^\circ$, consistent with the Kepler multi-planet baseline of $1.0$--$2.2^\circ$ \citep{Fabrycky2014}. The SE-CJ posterior is shifted to smaller values, although the two distributions overlap. We therefore state our result as a constraint on how much \emph{more} dispersed the SE-CJ systems could be, rather than as a detection of a smaller dispersion. Separate fits of SE-CJ and Non-SE-CJ samples yielded:
\begin{eqnarray*}
    \delta_{\rm SE-CJ}    & = & 0.86^{+0.47}_{-0.30}{}^\circ,
    \quad \kappa_{\rm SE-CJ} < 4.6\ (95\%), \\
    \delta_{\rm Non-SE-CJ} & = & 1.48^{+0.17}_{-0.15}{}^\circ,
    \quad \kappa_{\rm Non-SE-CJ} < 1.3\ (95\%).
\end{eqnarray*}

The $\kappa$ posteriors are statistically indistinguishable, motivating the shared-$\kappa$ model. Notably, these independent fits, in which $\kappa$ is unconstrained separately for each population, already show a factor-of-$\sim$1.7 difference in $\delta$ ($\delta_{\rm SE-CJ}=0.86^\circ$ vs.\ $\delta_{\rm Non-SE-CJ}=1.48^\circ$, with $P(\delta_{\rm SE-CJ}>\delta_{\rm Non-SE-CJ})=0.115$ treating the two fits as independent), confirming that this behaviour does not depend on sharing $\kappa$ between populations. The shared-$\kappa$ posteriors are:

\[
\begin{array}{cc}
\delta_{\rm SE-CJ} = 1.00^{+0.59}_{-0.37}{}^\circ &
\delta_{\rm Non-SE-CJ} = 1.48^{+0.17}_{-0.15}{}^\circ \\[0.6ex]
\multicolumn{2}{c}{\kappa < 1.3\ (95\%)} \\[0.6ex]
\multicolumn{2}{c}{P(\delta_{\rm SE-CJ}>\delta_{\rm Non-SE-CJ}) = 0.204.}
\end{array}
\]

We quote the shared-$\kappa$ values throughout, which are the more conservative of the two. Expressed as an effect size, the ratio of the two dispersions is $\delta_{\rm SE-CJ}/\delta_{\rm Non-SE-CJ}=0.68^{+0.40}_{-0.25}$ under the shared-$\kappa$ model, and $0.58^{+0.32}_{-0.21}$ for the separate fits. Because both populations are passed through the same forward model, a multiplicative bias common to both largely cancels in this ratio, making it a more robust summary than either dispersion on its own.







\begin{figure}[htb]
    \centering
    \includegraphics[width=\columnwidth]{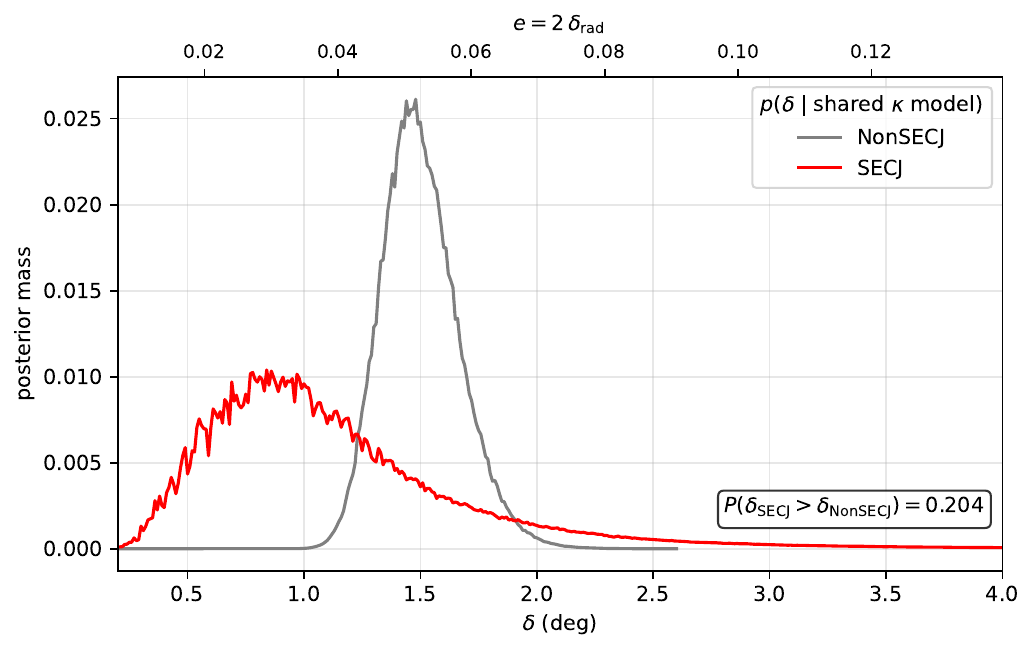}
    \caption{Marginal posteriors $p(\delta\mid\mathrm{data})$ for the
    mutual-inclination dispersion $\delta$ under the shared-$\kappa$ model,
    for SE-CJ systems (red) and Non-SE-CJ systems (gray). The top axis shows
    the corresponding eccentricity dispersion under equipartition assumption,
    $e\approx2\,\delta_{\rm rad}$.
    The posterior for Non-SE-CJ peaks near $\delta\approx1.5^\circ$
    ($e\approx0.05$), consistent with the Kepler multi-planet baseline model 
        of \citet{Fabrycky2014}. The SE-CJ posterior is shifted to smaller
    values ($\delta\approx1.0^\circ$, $e\approx0.035$), with
    $P(\delta_{\rm SE-CJ}>\delta_{\rm Non-SE-CJ})=0.204$ annotated. The two posteriors overlap substantially; we therefore report this as a constraint on excess dispersion in SE-CJ systems rather than as a detection of reduced dispersion.
    }
    \label{fig:delta_post}
\end{figure}


The equipartition assumption $e \approx 2\,\delta_{\rm rad}$ corresponds to a theoretical value $\kappa = 2$. We adopt it to express our $\delta$ values on an eccentricity scale comparable to previous work, not as an independent eccentricity measurement. Under this assumption, our measured mutual inclination dispersions correspond to eccentricity dispersions of $e\approx0.05$ for Non-SE-CJ and $e\approx0.035$ for SE-CJ systems.
Our own shared-$\kappa$ constraint is lower, $\kappa < 1.3$ at 95\% confidence, driven mainly by the much larger Non-SE-CJ sample. We note that the best-fitting model of \citet{Fabrycky2014} itself has $\sigma_e\simeq\sigma_i$ in radians, i.e.\ $\kappa\approx1$, so our upper limit agrees with that work even though it sits below equipartition; fitted separately, the SE-CJ sample alone gives $\kappa\approx1.9$. In either case $\kappa$ is poorly constrained by $\xi$ alone, and the quoted eccentricities should be read as indicative conversions.





\section{Discussion} \label{sec:discussion}


The inferred $\delta_{\rm Non-SE-CJ}=1.48^{+0.17}_{-0.15}{}^\circ$ falls
within the $1.0$--$2.2^\circ$ range that \citet{Fabrycky2014} find acceptable for
\textit{Kepler} multi-planet systems, validating our forward model and confirming that the Non-SE-CJ sample is representative of the broader Kepler multi-planet population. The corresponding equipartition eccentricity scale, $e\approx0.05$, is independently consistent with the photometric eccentricity measurements of \citet{Gilbert2025EccentricityValues}, who find $\langle e\rangle = 0.05\pm0.01$ via a methodologically distinct approach. The SE-CJ inner systems are found to be no more dispersed, and if anything flatter
($\delta_{\rm SE-CJ}=1.00^{+0.59}_{-0.37}{}^\circ$), consistent with a picture in which the inner SEs are strongly coupled with each other such that secular perturbations from the outer CJ are ineffective at exciting mutual inclinations between the inner planets (see Section~\ref{sec:disc_coupling} for details). Briefly, the strongly coupled inner planets undergo quasi-rigid nodal precession as a group, and the CJ's perturbations are shared across all inner planets rather than exciting large differential inclinations \citep{Rodet2021InclinedCJPertubation,Lai2017CJPerturbInclination}. The large semi-major axis separation between the inner SE chain (typically $\sim0.1\,\mathrm{AU}$) and the outer CJ ($\gtrsim1\,\mathrm{AU}$) is the primary reason for the weak SE-CJ secular coupling relative to the strong mutual SE-SE coupling within the compact inner system \citep{Lai2017CJPerturbInclination,Becker,Lammers_gap,Lubin}. 

In addition to secular coupling strength, the inclination of an outer CJ itself affects its ability to excite mutual inclinations between the SEs, with a CJ that is inclined doing so more effectively than a coplanar one \citep{Rodet2021InclinedCJPertubation}. \citet{Masuda2020CJinclinationPertubator} found that in systems with two or more transiting inner SEs, the mutual inclination between the CJ orbital plane and the inner SE plane is only $\sigma \approx 4^\circ$ i.e. favoring alignment with the inner system. This near-coplanarity is necessarily true by selection for transiting CJs, since their transit visibility already requires near-coplanarity with our line of sight. For the 18 SE-CJ systems in our sample whose CJs are detected only by radial velocity or transit-timing variations, no such transit-geometry constraint applies, and the CJ-SE relative inclination is in principle unbounded.  


Among the $N=22$ SE-CJ CJs in our sample with directly measured eccentricities (excluding 15 for which only a circular orbit was assumed in the published fit, and 2 for which only upper limits are available), the median is $\tilde{e}=0.09$ and the maximum is $e=0.40$, with $77\,\%$ below $e=0.2$, though individual systems can exceed this threshold (e.g.\ Kepler-65 with $e_{\rm c}\sim 0.30$); the SE-CJ distribution is significantly lower than the general field giant planet population ($\tilde{e}=0.20$, $49\,\%$ below $e=0.2$; two-sample KS $p=0.014$).\footnote{The field sample consists predominantly of RV-detected giants and is eccentricity-complete in a way our SE-CJ sample is not. Including the 15 assumed-circular entries in the SE-CJ sample would only strengthen the offset. Additionally, the field comparison includes giants around hosts of all metallicities and multiplicity classes, while SE-CJ systems are preferentially metal-rich; this metallicity difference may contribute modestly to the eccentricity offset independent of dynamical history.} In contrast, highly eccentric Jupiters ($e\gtrsim0.6$; e.g.\ TOI-3362b with $e=0.815$; \citealt{Dong2021HJTidalMigration}) are found predominantly in systems without detected inner companions. Together, these observations suggest that the CJs in SE-CJ multi-transiting systems did not undergo violent dynamical excitation. The flatness of both the inner SE planes (this work) and the CJs themselves \citep{Masuda2020CJinclinationPertubator} thus points to a dynamically cold evolution. Several known singly-transiting SE-CJ systems do host highly inclined and eccentric CJ companions (e.g., $\pi$ Men, HAT-P-11, WASP-107; \citealt{Xuan2020piMen}; \citealt{An2024HATP11}; \citealt{Piaulet}). Highly eccentric warm/hot Jupiters ($e\gtrsim0.6$) are found in systems lacking detected inner companions \citep{Dong2021HJTidalMigration}, consistent with the picture that strong planet-planet scattering or high-eccentricity migration, the mechanisms that excite such extreme eccentricities also stir up inner SE planets.

\subsection{Secular coupling strengths}
\label{sec:disc_coupling}

\begin{figure}
    \centering
    \includegraphics[width=\linewidth]{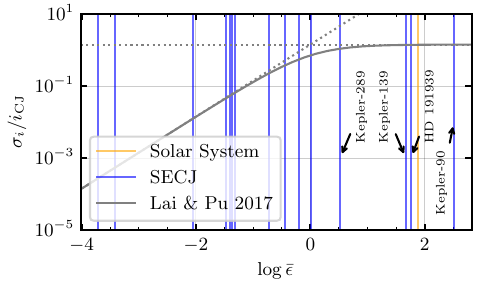}
    \caption{The distribution of secular coupling strengths for our 17 SE-CJ systems and the solar system and theoretical predictions of the resulting SE mutual inclination from \citet{Lai2017CJPerturbInclination}. The dashed gray lines show the limiting behavior for $\bar\epsilon\gg1$ (weak SE-SE coupling, strong SE-CJ coupling) and $\bar\epsilon< 1$ (strong SE-SE coupling, weak SE-CJ coupling). The solid gray line shows a smooth transition between the two that qualitatively resembles the results of numerical simulations with SEs of similar mass. 12/17 systems fall in the category where SE-SE secular coupling is stronger than SE-CJ coupling.}
    \label{fig:coupling}
\end{figure}

Adopting the methods presented in \citet{Lai2017CJPerturbInclination}, we estimated the strength of secular coupling between the SEs themselves and the secular coupling between the SEs and the CJs. This analysis is limited to systems containing multiple SEs. For systems with multiple CJs, we report the coupling with the innermost CJ: $m_{\text{CJ}}/a_{\text{CJ}}^3$. There are 17 SE-CJ systems amenable to this analysis all shown in Figure~\ref{fig:coupling}. For the SEs in these systems, we adopt the literature-reported masses; if not available, we estimated masses from their radii using the mass-radius relationship \texttt{spright} \citep{Parviainen2025Spright}, which is calibrated specifically for small planets and is therefore more accurate than \texttt{Forecaster} in this regime. For completeness, we included the SE-mass planets with $P>100$d in Kepler-139 and Kepler-90 as SEs for this analysis. 

For each of these systems, we compute the average secular coupling ratio with the CJ relative to the coupling between the SEs \citep{Lai2017CJPerturbInclination}
\begin{equation}
    \bar{\epsilon} = \left(\frac{1}{N-1}\sum_{j\neq d}\left|\epsilon_{jd}\right|^2\right)^{1/2},
\end{equation}
where $N$ is the number of SEs in the system, and $\epsilon_{jd}$ is the coupling strength between the $j$th SE and the dominant SE, here taken to be the SE with the largest mass. This coupling strength is given by \citep{Lai2017CJPerturbInclination}
\begin{eqnarray}
    \epsilon_{jd} &=&
    \left(\frac{m_{\text{CJ}}}{m_d}\right)
    \left(\frac{a_d}{a_{\text{CJ}}}\right)^3
    \left[\frac{3a_j/a_d}{b^{(1)}_{3/2}\left(a_j/a_d\right)}\right] \nonumber \\
    &&{}\times
    \frac{\left(a_d/a_j\right)^{3/2}-1}
    {1 + \left(m_j/m_d\right)\left(a_j/a_d\right)^{1/2}} .
\end{eqnarray}
where $b^{(1)}_{3/2}(\alpha)$ is a Laplace coefficient\footnote{The Laplace coefficient $b^{(1)}_{3/2}$ is unrelated to the transit impact parameter $b$.} \citep{Murray1999SolarSystemDynamics}. The results of these calculations are shown in Figure~\ref{fig:coupling}. We note that $\epsilon_{jd}\propto m_{\text{CJ}}/a_{\text{CJ}}^3$, motivating our criterion for selecting the innermost CJ in multi-CJ systems. Because the cold Jupiter enters the coupling only through this factor, the fractional contribution of any additional outer companion can be evaluated exactly. Eight systems in our sample host more than one CJ; for six of them the outer companions add less than $2\%$ to $\bar\epsilon$, and at most $32\%$ for the other two. These corrections are negligible against the four orders of magnitude spanned by $\bar\epsilon$ across our sample, and leave every system in the same coupling regime. We also repeated this analysis on the Solar System for comparison, taking the Earth to be the dominant ``SE" and Jupiter as the primary CJ.

In the strong SE-SE coupling limit, corresponding to $\bar\epsilon < 1$, the CJ's ability to excite mutual inclination among the SEs is limited to $\sigma\sim \bar\epsilon i_{\text{CJ}}$ \citep{Lai2017CJPerturbInclination}. The CJs in these systems are unable to induce significant mutual inclinations among the SEs. See also the simulations by \citet{Gratia} and \citet{Lammers_gap}. 12/17 of our SE-CJ have $\bar\epsilon < 1$. 

When $\bar\epsilon \sim 1$, there is the possibility for secular resonances to induce high relative inclination between SE's within the system \citep{Lai2017CJPerturbInclination}. However, these resonances are only significant when there are large disparities in the SE masses within the SE-CJ system (i.e. one SE's mass being $\gtrsim5\times$ that of another).

In the case of weak coupling between the SEs, $\bar\epsilon \gg 1$, the CJ is able to strongly influence the SEs in the system, and excite mutual inclination comparable to its own inclination. This is the case in the Solar System, for which $\bar\epsilon\approx77$. Indeed, a Laplace-Lagrange model of the Solar System displays long-term inclination variation driven by the giant planets \citep{Murray1999SolarSystemDynamics}. Four systems from our sample also have $\bar\epsilon \gg 1$: Kepler-289 ($\bar\epsilon\approx3$), HD 191939 ($\bar\epsilon\approx 57$), Kepler-139 ($\bar\epsilon\approx46$), and Kepler-90 ($\bar\epsilon\approx325$). Although this limit suggests that the CJs in these four systems could induce significant mutual inclination among the SEs, this excitation is limited by the CJ's own inclination. See also, simulations by \citet{Livesey} who put the CJ close to the inner SE (at 1.5--2.0 semi-major axis ratio). In this case, the CJ has strong coupling with the SEs, and indeed \citet{Livesey} found that SE mutual inclinations can be excited.

Taking this analysis as a whole, the SEs in the great majority of our SE-CJ sample (12/17 systems with $\bar\epsilon < 1$) are coupled with each other much more strongly than with the outer CJ. In other words, in their typical configuration, CJs at $\sim 1$~AU are simply too distant or too low in mass to dynamically excite mutual inclinations among inner SEs at $\sim 0.1$~AU. This is the dynamical picture our non-detection of elevated $\delta_{\rm SE-CJ}$ result is most consistent with.

Our treatment is purely secular and would not apply to pairs locked in, or close to, a mean-motion resonance; a future study including both secular and resonant terms may be warranted. If the SEs are locked in librating mean motion resonances, they are even more resistant to secular perturbation from a cold Jupiter (Huang et al. in preparation). We also acknowledge that this section is based on a Laplace-Lagrange model of secular perturbation; the conclusion could be modified if secular chaos is important \citep[e.g.][]{Lithwick2011}.

\subsection{Gap Complexity?}
\label{sec:disc_heweiss}

\citet{HeWeiss2023GapComplexity} found that the four systems in the Kepler Giant Planet Search catalog with $\geq3$ inner transiting small planets and an outer giant companion (Kepler-48, Kepler-65, Kepler-90, and Kepler-139) have substantially higher gap complexity \citep{Gilbert2020SystemArchitectures} than systems lacking outer giants, and attributed this to dynamical influence from the outer giant. All four of these systems are included in our transit duration ratio analysis.


Gap complexity of a transiting planetary system can be produced by both orbital spacing irregularity as well as non-transiting planets with large mutual inclinations. The measured orbital inclinations of the transiting planets in the four systems reported by \citet{HeWeiss2023GapComplexity} suggest low mutual inclinations: computing the (unweighted) standard deviation of the orbital inclinations assuming the planets transit the same stellar hemisphere  yields $\sigma_i = 0.26^\circ$, $1.42^\circ$, $0.27^\circ$, and $1.17^\circ$ for Kepler-48, Kepler-65, Kepler-90, and Kepler-139, respectively. The overall dispersion is $0.83^\circ$ across all 17 transiting planets in the four systems. This is in good agreement with our forward-model inference of $\delta_{\rm SE-CJ} = 1.00^{+0.59}_{-0.37}{}^\circ$. The true mutual-inclination dispersion in these systems can only substantially exceed this transiting-planet estimate if each system harbors a large population of non-transiting planets that are systematically more inclined; we consider this unlikely.


All four of these systems have SE-CJ coupling strength estimates in Section~\ref{sec:disc_coupling}. Kepler-48 and Kepler-65 have weak SE-CJ coupling ($\bar\epsilon\sim0.3$ and $\bar\epsilon\sim0.04$, respectively) that is unable to excite strong mutual inclinations. \citet{Gilbert2026TheOtherPaper} also found no statistically significant association between eccentricity and gap complexity across the broader Kepler compact multi population, suggesting that spacing irregularity and dynamical temperature are largely decoupled. We tentatively interpret the elevated gap complexity of these two systems as a formation or migration fossil. In metal-rich systems (both Kepler-48 and Kepler-65 have super-solar metallicity: 0.242$\pm$0.042 and 0.122$\pm$0.043; \citealt{Petigura2017CKS_I}) that form both CJs and inner SEs, the innermost SE may form quickly and Type-I migrate to the inner disk edge ($\sim0.05\,$AU, \citealt{Masset}) leaving an apparent gap \citep[]{Millholland_gap} with the outer CJs which undergo slower Type-II migration \citep{Kley}. The resultant gap in period distribution (gap complexity) is thus an imprint of this formation history rather than dynamical excitation.

Kepler-90 and Kepler-139, in contrast, show strong coupling between the SEs and CJs ($\bar\epsilon\sim325$ and $\bar\epsilon\sim46$, respectively). In other words, any inclination of the CJ would be transferred effectively to the inner SEs. However, it is still worth noting that the amount of mutual inclination that can be excited is limited by the initial inclination of the CJ. At the proposed gaps in Kepler-90 and  Kepler-139 (Figure \ref{Results.Architecture.Image}), a small mutual inclination of 1--2$^\circ$ is more than enough to produce a non-transiting geometry for a planet that resides in the gap. In short, even with the observed gap complexity in Kepler-90 and Kepler-139, the overall orbital architectures are likely close to being coplanar. \citet{Lammers_gap} performed a more detailed analysis than ours. They inserted a hypothetical planet within the putative period-ratio gap and quantified its secular coupling to the neighboring planets. They found through direct simulations that for all of the four systems mentioned in \citet{HeWeiss2023GapComplexity}, mutual-inclination excitation was not observed. \citet{Lammers} analysed Kepler-139 in detail, combining transit timing with radial velocities to identify a non-transiting planet, Kepler-139\,f, orbiting interior to the outer giant. Their secular integrations show the giant acting on the outer edge of the inner system: planet f is driven out of a transiting geometry, while the innermost pair, Kepler-139\,d and b, remains mutually aligned to well below a degree. Kepler-139 is therefore a system in which the cold Jupiter does measurably influence the inner architecture, as its position in Figure~\ref{fig:coupling} would lead one to expect, but in which that influence is confined to the planets least strongly coupled to the rest of the chain. The planets we observe transiting remain consistent with a near-coplanar configuration.

\subsection{Caveats}
\label{sec:disc_survivor}

Transit-duration-ratio inference is feasible only for multi-transiting systems. Because systems with large mutual inclinations are intrinsically less likely to exhibit multiple transiting planets, the SE-CJ systems analyzed here represent a biased subset: the most strongly perturbed SE-CJ architectures are more likely to appear as single-transiting systems and are therefore excluded from our analysis. This selection effect is consistent with predictions from N-body scattering experiments \citep{Huang2017hotSEfromScattering}, which show that single-transiting SEs with outer giant companions are systematically more eccentric and mutually inclined than their multi-transiting counterparts, as exemplified by $\pi$ Mensae \citep{Xuan2020piMen}. Our inferred $\delta_{\rm SE-CJ}$ is therefore likely a lower limit on the true dispersion of the broader SE-CJ population.

We note that this selection effect applies symmetrically to both populations in the $\xi$ analysis: because computing $\xi$ requires an adjacent transiting pair, the Non-SE-CJ comparison sample is, by construction, also restricted to multi-transiting systems. The relative contrast we measure between SE-CJ and Non-SE-CJ therefore remains meaningful even though the absolute $\delta$ values for both populations are lower bounds on their respective full (single- and multi-transiting) populations.

\citet{Hua} found higher mutual inclinations for the shortest-period Kepler-like planets in higher-metallicity systems. This trend seems to be driven by the inclusion of ultra-short-period planets whose inward migration also excited mutual inclinations \citep{Dai2018}. \citet{Petrovich2019} and \citet{Pu2019} attributed those mutual inclination excitations to the inner SEs themselves.
We also verified that the metallicity offset between the SE-CJ and Non-SE-CJ samples \citep{Zhu2024SECJmetallicity} does not drive the $\xi$ result. Within the Non-SE-CJ sample, metal-rich hosts ($\mathrm{[Fe/H]}>0$) have statistically lower $T/T_0$ than metal-poor hosts (KS $p=0.001$), consistent with the finding by \citet{Hua} that metal-rich Kepler systems have higher mutual inclinations. Were metallicity the explanation for the SE-CJ-Non-SE-CJ $\delta$ difference, the metal-rich SE-CJ sample should appear \emph{more} dynamically excited than the average Non-SE-CJ; the opposite is observed. The metallicity offset thus works against the observed SE-CJ flatness, strengthening rather than weakening our conclusion that SE-CJ systems are not more mutually inclined than Non-SE-CJ systems. This is consistent with \citet{Gilbert2026TheOtherPaper}, who found no significant association between host metallicity and eccentricity across the broader Kepler compact multi-planet population.


Our forward model adopts a single-component Rayleigh distribution for mutual inclinations, which may not fully capture the true diversity of
SE-CJ architectures.  Additionally, we do not propagate per-planet duration measurement
uncertainties. We also treated all pairs within a system as independent. Our samples are only mildly clustered ($1.3$ pairs per system for SE-CJ, $1.5$ for Non-SE-CJ); a system-level block bootstrap, resampling whole systems rather than individual pairs, widens the posteriors by factors of $1.02$ and $1.12$ respectively, so the quoted intervals are optimistic at roughly the ten per cent level, and assume a uniform $\omega$ distribution. These assumptions can be relaxed in a future study when a larger sample is available.

Because long-period giants are detected primarily via radial velocity, and the Non-SE-CJ sample is defined by the absence of a detected CJ, some
fraction of the Non-SE-CJ systems likely harbor undetected outer giants. This contamination dilutes the SE-CJ-to-Non-SE-CJ contrast, and the true
difference in $\delta$ between CJ-hosting and CJ-free systems may be larger than what we measure.
We attempted to construct an RV-vetted Non-SE-CJ subsample to test this
directly. Long-baseline, high-precision radial velocities capable of excluding a
Jupiter-mass companion near $1$~AU exist for only the 63 Kepler systems of the
Kepler Giant Planet Search \citep{Weiss}; of the 295 systems
contributing to our Non-SE-CJ sample, 27 have such coverage. These are not a
representative subset: because precise radial velocities require bright hosts,
their transits have nearly twice the median signal-to-noise of the remaining
Non-SE-CJ systems (KS $p\sim10^{-10}$) and a higher median transit multiplicity.
Both differences bias the inferred $\delta$ downward independently of whether a
cold Jupiter is present, and they exceed the contamination effect the comparison
is intended to isolate. We therefore do not report a vetted-subsample
$\delta_{\rm Non-SE-CJ}$. This contamination acts to dilute rather than to create
the reported contrast. Targeted long-baseline RV monitoring of the Non-SE-CJ hosts, together with future PLATO discoveries \citep{plato} of transiting systems around brighter and closer hosts, would substantially improve both sample purity and size. We note that Gaia astrometry \citep{gaia}, while powerful for nearby stars, cannot meaningfully constrain cold Jupiters at Kepler distances: a Jupiter-mass companion at $1$~AU around a solar-mass host at $800$~pc produces an astrometric signal of order $1\,\mu$as, well below Gaia's per-epoch precision.




\section{Conclusions} \label{sec:conclusions}

We compared the orbital architectures of Super-Earth--Cold-Jupiter (SE-CJ) systems with the broader Kepler-like systems (Non-SE-CJ), using two transit-duration-based diagnostics:

\begin{enumerate}

\item The normalized transit duration $T/T_0$ shows no statistically significant difference between SE-CJ and Non-SE-CJ Super-Earth populations
(KS test $p=0.20$). This does not support the claim that SE-CJ systems are more dynamically excited than Non-SE-CJ systems. 

\item Forward modeling of pairwise transit duration ratios $\xi$ for adjacent SE-SE pairs shows no evidence that SE-CJ systems are more mutually inclined than Non-SE-CJ systems:
$\delta_{\rm SE-CJ}=1.00^{+0.59}_{-0.37}{}^\circ$ versus $\delta_{\rm Non-SE-CJ}=1.48^{+0.17}_{-0.15}{}^\circ$, with $P(\delta_{\rm SE-CJ}>\delta_{\rm Non-SE-CJ})=0.20$. The inferred dispersions mildly favor the SE-CJ systems being flatter, $\delta_{\rm SE-CJ}/\delta_{\rm Non-SE-CJ}=0.68^{+0.40}_{-0.25}$, and this preference strengthens to $P(\delta_{\rm SE-CJ}>\delta_{\rm Non-SE-CJ})=0.06$ when the single discrepant system Kepler-139 is excluded (Section~\ref{sec:likelihood}), but it does not reach statistical significance in our adopted sample. Under the equipartition assumption, these correspond to orbital eccentricity of $e\approx0.035$ and $e\approx0.05$, respectively. The Non-SE-CJ result is consistent with the Kepler multi-planet baseline of \citet{Fabrycky2014} and with the eccentricity measurements of \citet{Gilbert2025EccentricityValues}.

\end{enumerate}
Two caveats: first, our transit-duration-ratio analysis is feasible only for multi-transiting systems; SE-CJ systems in which the inner planets were strongly excited by the outer giant are expected to present as single-transiting and therefore fall outside the scope of this analysis, so our inferred $\delta_{\rm SE-CJ}$ likely underestimates the inclination dispersion of the broader SE-CJ population. Second, because the Non-SE-CJ sample is defined only by the absence of a currently detected CJ, some of these systems likely harbor undetected outer giants; this would dilute the measured contrast, and the true difference in $\delta$ between CJ-hosting and CJ-free systems may be larger than reported here.

Our calculations of secular coupling strengths \citep{Lai2017CJPerturbInclination} indicate that the SEs are coupled with each other more strongly than with the CJs. In other words, CJs ($\sim 1$~AU) are often too far or too low in mass to dynamically excite mutual inclinations among the inner SEs ($\sim 0.1$~AU).

Our results support a picture in which many SE-CJ inner systems are dynamically cold, particularly those with multi-transiting SEs that are amenable to our transit-duration-ratio analysis. We tentatively suggest that this may reflect either coplanarity between the outer CJ and the inner SE system born out of a quiescent formation channel, or a natural separation in orbital periods produced by fast Type-I migration of SEs and  slow Type-II migration of CJs, which would weaken secular coupling between the two populations.

\begin{acknowledgments}\label{acknowledgments}
We thank Dan Fabrycky, Ji-Wei Xie, Wei Zhu, Quanyi Liu, Lauren Weiss, Marco Scharringhausen, Caleb Lammers, and Joshua Winn for helpful discussion and detailed comments on an early draft. This research has made use of the Exoplanet Follow-up Observation Program (ExoFOP; DOI: 10.26134/ExoFOP5) website, which is operated by the California Institute of Technology, under contract with the National Aeronautics and Space Administration under the Exoplanet Exploration Program.\\
\end{acknowledgments}

\software{numpy \citep{Harris2020Numpy}, pandas \citep{McKinney2011Pandas}, 
matplotlib \citep{Hunter2007Matplotlib}, SciPy \citep{Virtanen2020SciPy}, 
Forecaster \citep{Chen2016Forecaster}, spright \citep{Parviainen2025Spright}}

\bibliographystyle{aasjournal}
\bibliography{bibliography}

\end{document}